\documentclass[aps,pre,onecolumn,showpacs,amsmath,amssymb,longbibliography,notitlepage,11pt,tightenlines]{revtex4-1} 
\usepackage{graphicx} 
\usepackage{dcolumn} 
\usepackage{bm} 

\usepackage{xcolor}

\DeclareMathOperator\erfc{erfc}
\newcommand{\fmov}{f_\text{act}}
\newcommand{\fmovn}{f_\text{act,n}}
\newcommand{\geff}{\gamma_{\text{eff}}}

\begin{document}

\title{A model for approximately stretched-exponential relaxation with continuously varying stretching exponents}

\author{Joseph D. Paulsen}
\email{jdpaulse@syr.edu}
\affiliation{Department of Physics, Syracuse University, Syracuse, NY 13244, USA}
\author{Sidney R. Nagel}
\affiliation{The James Franck and Enrico Fermi Institutes and The Department of Physics, The University of Chicago, Chicago, IL 60637, USA}

\date{\today}

\begin{abstract}
Relaxation in glasses is often approximated by a stretched-exponential form: $f(t) = A \exp [-(t/\tau)^{\beta}]$. 
Here, we show that the relaxation in a model of sheared non-Brownian suspensions developed by Cort{\'e} \textit{et al.}~[Nature Phys.~{\bf 4}, 420 (2008)] can be well approximated by a stretched exponential with an exponent $\beta$ that depends on the strain amplitude: $0.25 < \beta < 1$. 
In a one-dimensional version of the model, we show how the relaxation originates from density fluctuations in the initial particle configurations. 
Our analysis is in good agreement with numerical simulations and reveals a functional form for the relaxation that is distinct from, but well approximated by, a stretched-exponential function.
\end{abstract}

\maketitle

\section{Introduction}

The relaxations of out-of-equilibrium disordered systems can be strongly non-exponential. 
In many cases the time-dependence of relaxation follows the Kohlrausch-Williams-Watts stretched-exponential function: 
\begin{equation}
f(t) = A \exp[-(t/\tau)^{\beta}],
\label{strexp}
\end{equation}

\noindent where the stretching exponent $0<\beta<1$ parameterizes the degree of departure from a simple exponential. 
This functional form has been used to describe the mechanical, electric, and magnetic response of a remarkably broad range of materials, including structural glasses and polymers~\cite{Williams70,Legrand87,Ediger96}, spin and magnetic glasses~\cite{Chamberlin84,Coey87}, charge- and spin-density wave carriers~\cite{Kriza86,Mihaly91}, amorphous silicon~\cite{Kakalios87}, and a crumpled sheet of mylar~\cite{Kramer96}. 

Numerous models of the transport and trapping of electrons or defects have been constructed to deduce this mathematical form~\cite{Grassberger82,Klafter86,Palmer84,Scher91,Phillips96,Jund01,Sturman03,Lemke11}. 
However, few models address the spatial structure of crowded particles, which is important for understanding the non-exponential mechanical response of disordered particulate media such as structural glasses and soft materials, including foams, colloids, and granular matter~\cite{Knight95,Ono03,Mattsson09,Zou10}. 

Here we study glassy relaxation in a simplified computer simulation model of cyclically sheared particles, which we analyze in real-space. 
The simplest version of the model that we study is the antithesis of complexity: particles are confined to one dimension, where they interact repulsively, and only with their two neighbors. 
We show how a wide distribution of timescales and approximate stretched-exponential relaxation can arise from these simple ingredients. 
In addition, by making the interactions only slightly more complex (i.e., by allowing attractive or repulsive interactions with equal probability), we show that the stretching exponent, $\beta$, can be varied between $0.25$ and $1$ by changing the driving amplitude. 

The basic mechanism underlying the behavior of both model variants is simple and robust: particles that are crowded, due to random initial conditions, must get out of each other's way in order to relax. 
In the models we study, density fluctuations occur on a wide range of scales due to the initial conditions -- such disorder would be present in a glass. 
Because longer-wavelength crowding requires more time to expand, the end result is a broad distribution of relaxation timescales.

\textit{Relaxation of sheared suspensions.}---
Our model is based on simulations of a non-Brownian viscous suspension under cyclic shear. 
These simulations were introduced by Cort\'e \textit{et al.}~\cite{Corte08} to explain irreversibility and self-organization in particle suspensions, and they display a wealth of phenomena, including a dynamical phase transition for strains smaller than a critical amplitude~\cite{Corte08,Menon09,Xu13}. 
Qualititative analogies with glassy systems were recently found near this critical point, such as spatial correlations and diverging timescales~\cite{Tjhung15,Tjhung16}.
There are expected to be extremely homogeneous spatial distributions near the critical point~\cite{Hexner15,Tjhung15}, which were observed in a closely related experiment~\cite{Weijs15}.
The model was also found to support multiple transient memory effects~\cite{Keim11,Keim13,Paulsen14}. 
These transient memories are an analog in fluids of the multiple transient memories found earlier in charge-density-wave conductors~\cite{Coppersmith97,Povinelli99}. 
Here, we study the relaxation dynamics.

\begin{figure}[bt]
\centering 
\begin{center} 
\includegraphics[width=\textwidth]{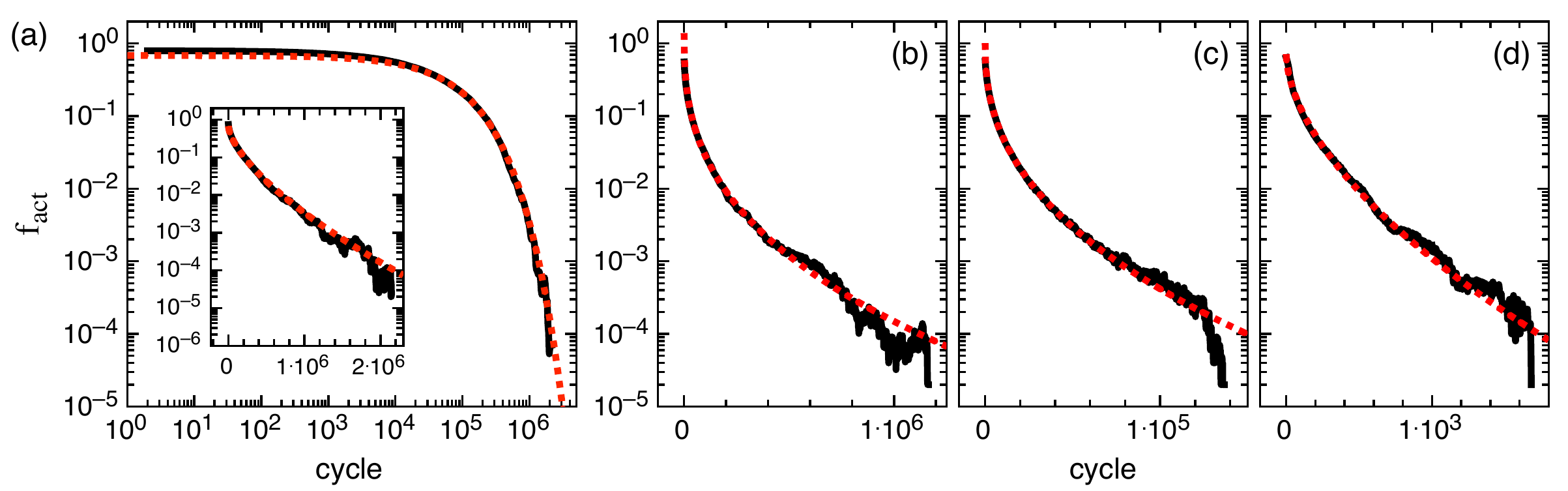} 
\end{center}
\caption{
{\bf Fraction of active particles, $\fmov$, versus cycle for different particle interaction rules.} 
{\bf (a)} Two-dimensional model (see text) for a single system of $N=10^5$ particles with density $\phi=0.2$, strain amplitude $\gamma=3$, and maximum kick size $\epsilon=0.01$. 
Main figure shows results on log-log axes. 
Inset shows results on linear-log axes. 
Dotted red line: stretched-exponential fit (eqn.~\ref{strexp}) with $A=0.68$, $\tau=77000$, and $\beta=0.65$. 
{\bf (b-d)} One-dimensional models in which only nearest-neighbors interact with various collision rules. 
Swelling amplitude is $\gamma = 0.5$. 
Dotted red lines show fits to eqn.~\ref{strexp} with stretching exponents of $\beta=$ 0.37, 0.42, and 0.57, respectively. 
{\bf (b)} Overlapping particles each receive a random magnitude kick in a random direction. 
{\bf (c)} ``Neutral model" where kicks have a fixed magnitude of 0.01 and can be either attractive of repulsive. 
The kicks between pairs of particles are in opposite directions and thus conserve center of mass. 
{\bf (d)} ``Purely-repulsive model", the same as (c), but where kicks are always repulsive. 
}
\label{Fig1}
\end{figure}

In the simulation, $N$ particles of diameter $d=1$ are distributed randomly in a box of area $A_{\text{box}}$ at area fraction $\phi=N\pi /4A_\text{box}$ with periodic boundary conditions. 
An affine shearing deformation is applied to the particle centers, with a strain amplitude that grows continuously from $0$ to $\gamma$. 
The particles are then returned to the initial configuration at zero strain. 
In this simplified picture of a sheared suspension, particles are allowed to overlap and pass through one another. 
Two particles are said to ``collide" if they overlap at any point in the cycle. 
To simulate the effect of collisions, each colliding particle is given a small displacement in a random direction at the end of the cycle. 
During a given cycle, a particle is called ``active" if it collides with at least one other particle (and it gets multiple kicks if it overlaps with multiple particles). 
These displacements have a random magnitude between $0$ and $\epsilon$. 
Except where otherwise stated, we set $\epsilon=0.01$ and $N=10^5$. 
(The qualitative behavior is independent of these values.) 
Figure \ref{Fig1} shows the fraction of active particles, $\fmov$, versus cycle number. 
The data follow an approximately stretched-exponential form with $\beta \approx 0.65$. 

\section{Model variants studied}

We wish to uncover the minimal features of this two-dimensional model that produce the approximate stretched-exponential decay. 
To this end, we consider a simplification of this model to one dimension, following ref.~\cite{Corte08}, where $N$ point particles are randomly distributed on a line of length $N$ and are swelled cyclically to diameter $0<\gamma<1$. 
After each cycle, particles are displaced by a random amount between $-\epsilon$ and $+\epsilon$ for each collision with another particle. 
Figure~\ref{Fig1}b shows that in this one-dimensional model, $\fmov$ versus cycle number has a stretched-exponential form ($\beta \approx 0.37$). 

The remainder of this paper is devoted to two even simpler variants of this one-dimensional model, which we study in a range of amplitudes where the particles always reach a quiescent state (i.e., $\gamma<\gamma_{\text{c}}$ \cite{Corte08}).
In both variants, particles interact in one dimension with only their two nearest neighbors. 
In both cases, the models exhibit approximately stretched-exponential relaxation. 

In the ``neutral model" (Fig.~\ref{Fig1}c and section~\ref{neutral}), the kicks are all of the same magnitude $\epsilon$ and in opposite directions for the two colliding particles.  
The kicks are either attractive (bringing the particles closer together) or repulsive (pushing them apart) with equal probability. 
Thus, the center of mass of the particles is conserved during each collision. 

In the ``purely-repulsive model" (Fig.~\ref{Fig1}d and section~\ref{repulsive}), the collisions again preserve the center of mass but are always repulsive with fixed size $\epsilon$. 
We emphasize that this interaction rule produces fully deterministic dynamics: the initial particle positions exactly specify the evolution of the system.

\section{Approach}

To understand these dynamics, we first consider the structure of the final state. 
For both models, in the limit of $\epsilon \rightarrow 0$, the final configuration consists of \textit{clusters} of particles spaced apart by exactly $\gamma$, with larger gaps separating the clusters from one another. 
Our approach will be to address the relaxation of the particles that will end up in a cluster of final size $n$. 
At any point in time, we denote the fraction of active particles among this population as $\fmovn(t)$. 
Then we can recover the total relaxation by summing over all these separate populations: 
\begin{equation}
\fmov(t) = \sum_{n=1}^{N} \rho(n,\gamma) n \fmovn(t), 
\label{fmovsum}
\end{equation} 
\noindent where $\rho(n,\gamma)$ is the fraction of final clusters having $n$ particles when the swelling diameter is $\gamma$. 

We will consider both the purely-repulsive case (Fig.~\ref{Fig1}d) and the repulsive-attractive case (Fig.~\ref{Fig1}c). 
We note that the final cluster size distribution, $\rho(n,\gamma)$, depends only on the initial positions and not on the dynamics for these two cases, since their collision rules conserve center of mass. 
This is because any group of overlapping particles will conserve its center of mass as it expands, and when two neighboring groups merge into a larger group, particles then expand about the new center of mass of the combined group. 

We begin by deriving this distribution, $\rho(n,\gamma)$, since it is the same in both models. 
We will then determine the relaxation timescale and form of the $\fmovn(t)$, which differ in the two models. 
We can then obtain expressions for $\fmov (t)$ by plugging into eqn.~\ref{fmovsum}. 

\begin{figure}[bt]
\centering 
\begin{center} 
\includegraphics[width=0.75\textwidth]{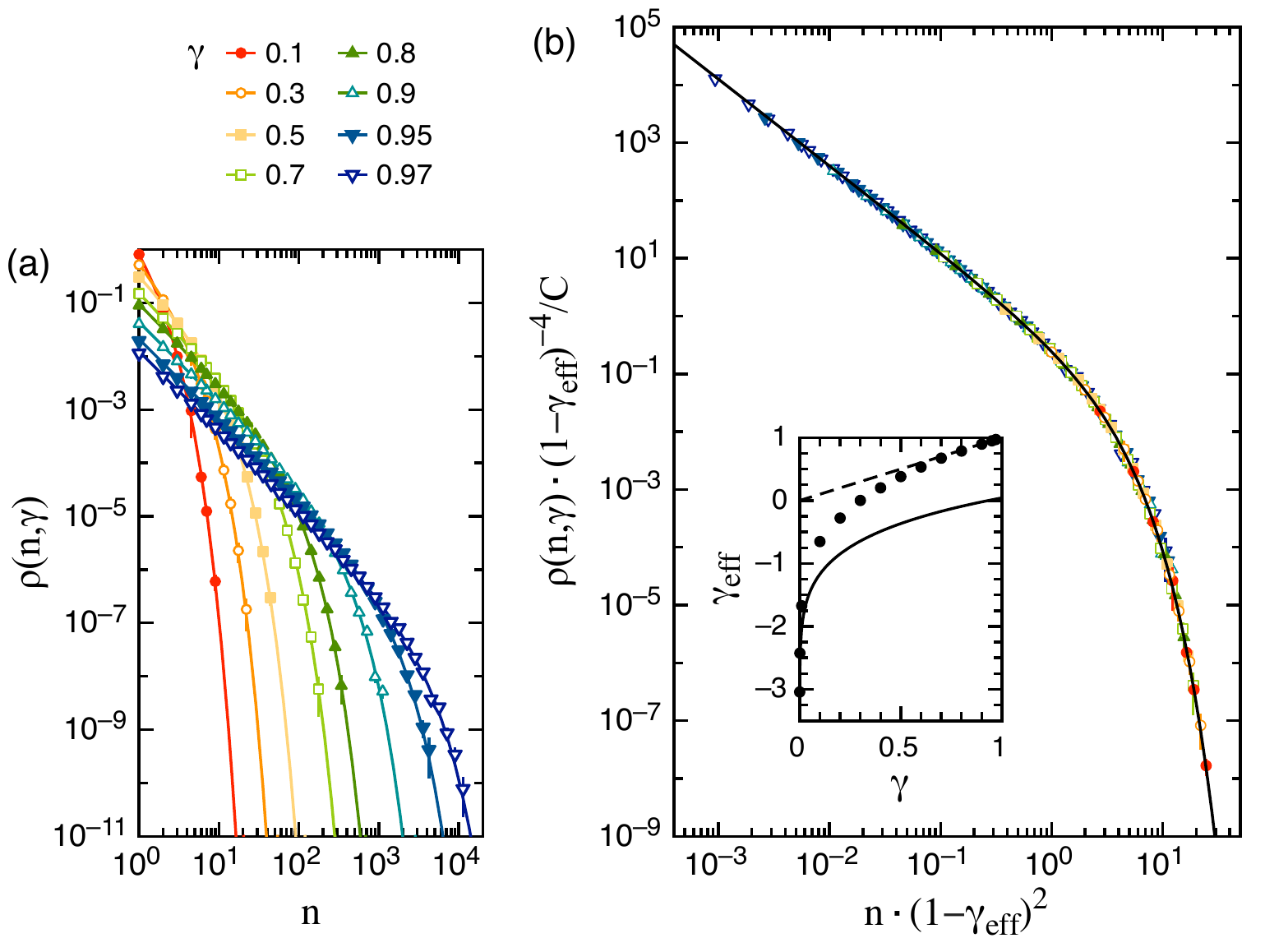} 
\end{center}
\caption{
{\bf Final cluster size distribution for one-dimensional models that conserve center of mass.} 
{\bf (a)} Distribution $\rho(n,\gamma)$ versus cluster size, $n$, for $\gamma$ ranging from $0.1$ to $0.97$ (using $N=10^7$ particles). 
The data are well described by eqn.\ \ref{rhon} (solid lines) if $\gamma$ is treated as a fitting parameter. 
{\bf (b)} The data collapse onto a master curve when the axes are rescaled. 
Solid line: eqn.\ \ref{rhon}. 
Inset: values from the fits, $\geff$, versus $\gamma$. 
For $\gamma$ near 1, the data approach $\geff=\gamma$ (dashed line). 
For small $\gamma$, the data approach $\geff=1-\sqrt{-2\ln[1-\exp(-\gamma)]}$ (solid line). 
}
\label{Fig2}
\end{figure}

\textit{Cluster size distribution.}---
Figure~\ref{Fig2}a shows measurements of the distribution of cluster sizes in the limit of small $\epsilon$. 
The data have a power-law form with a cutoff that depends on $\gamma$; larger clusters are produced for larger $\gamma$. 

Here we show how density fluctuations in the initial (random) configuration of particles lead to this final cluster distribution. 
These density fluctuations create regions of overlapping particles in the initial configuration that have two possible fates: either they expand in isolation if they are sufficiently far away from other particles, or they merge with a neighboring group, which can in turn expand and merge with another group. 
Thus, small overlapping groups of particles should only be counted as small clusters if they never get incorporated into a larger group. 
Therefore, to calculate the size of the cluster in which a particle will eventually reside, one must find the largest group of particles to which it could belong. 
The procedure is then to consider long-wavelength density variations first, and then look on smaller and smaller scales, until a set of particles is identified that will join together by the end of the simulation. 

We label the initial positions by $\{ x_i \}$ from left to right on the interval $[0,N]$, choosing the origin so that the leftmost particle of one of the final clusters is at $x_0=0$. 
To form a cluster of \textit{at least} size $n$, there must be $n$ particles within an interval of length $\gamma (n-1)$ in the initial configuration. 
Thus, we seek the largest $i$ for which $x_i \leq \gamma i$. 

To estimate the $x_i$, we note that the gaps between neighboring particles, $\Delta_i = x_{i+1} - x_i$, are random variables drawn from an exponential distribution. 
Thus, we can model the $x_i$ as steps on a random walk that starts at $x=0$ and ends at $x=N$.  

Approximating the $x_i$ and $i$ as being continuous, this problem can be mapped onto to a first-crossing problem with a known solution~\cite{Beghin99}. 
In particular, changing variables to $\bar{x} = -x + \gamma i + N(1-\gamma)$ and $t=N-i$, we seek the first crossing of the constant boundary $N(1-\gamma)$ by a Brownian bridge $\bar{x}(t)$ with $\bar{x}(0) = 0$ and $\bar{x}(N) = N(1-\gamma)$. 
This mapping yields the probability distribution of the first crossing occurring at $i=n$, which corresponds to the particle at the far left of the system belonging to a cluster of size $n$. 
Since there are $n$ particles in that cluster, we scale this result by $1/n$ in order to obtain the distribution of cluster sizes: 
\begin{equation} 
\rho(n,\gamma) = C\frac{(1-\gamma)}{\sqrt{2\pi} } \frac{\exp(-\frac{1}{2}n(1-\gamma)^2)}{n^{3/2}}, 
\label{rhon} 
\end{equation} 
\noindent where $C=C(\gamma)$ is determined by the normalization condition $\sum_{n=1}^N \rho(n,\gamma)n=1$, for each swelling size $\gamma$. 

We note that this mapping to a Brownian bridge is only valid when the cluster size, $n$, can be approximated as a continuous variable. 
This can be done in the dual limit $\gamma \rightarrow 1$ and $N \rightarrow \infty$. 
For smaller $\gamma$, the discreteness of $n$ introduces an error so that this expression in no longer exact. 

We can also solve for $\rho(n,\gamma)$ in the limit of small $\gamma$ by observing that in this case, the motions of the particles during relaxation will be negligible. 
Thus, $\rho(n,\gamma)$ can be approximated by the initial distribution of overlapping groups of particles. 
To derive this distribution, we consider the gaps between $N$ points that are randomly placed on a line of length $N$. 
The probability that the distance $\Delta$ between two adjacent points is larger than $\gamma$ is given by $P(\Delta>\gamma) = \exp(-\gamma)$. 
Thus, the probability that $n$ particles of diameter $\gamma$ form a contiguous chain is proportional to $[1-\exp(-\gamma)]^{n-1}$, which can be expressed as an exponential in $n$ with a decay constant of $\ln[1-\exp(-\gamma)]$. 

Figure~\ref{Fig2} shows that we can describe the entire range of data using eqn.~\ref{rhon}, if we treat $\gamma$ as a fitting parameter, $\geff$. 
We find $\geff \approx \gamma$ near $\gamma = 1$, indicating good agreement with the first-passage model. 
As $\gamma$ decreases, the $\geff$ depart from $\gamma$. 
We can describe the limit of small $\gamma$ by matching the decay constant in the exponential in eqn.~\ref{rhon} to $\ln[1-\exp(-\gamma)]$ as argued above; namely $\geff=1-\sqrt{-2\ln[1-\exp(-\gamma)]}$. 
Thus, we can interpolate between these two limiting cases for intermediate $\gamma$. 
Figure~\ref{Fig2} shows that this fitting form, which is exact in the two limits, gives an excellent description of the data for all relevant $\gamma$. 

\textit{Connection to slip avalanches.}---
The distribution of cluster sizes in these one-dimensional swelling models has a parallel to slip avalanches, which occur during the deformation of amorphous materials~\cite{Denisov16}. 
In particular, a simple model of slip avalanches~\cite{Dahmen09,Dahmen11} has a basic mechanism that is shared with the swelling model: small active regions can trigger other regions to become active, thus forming a chain reaction that leads to a single large avalanche. 
The scaling and cutoff of avalanche size in the mean-field version of the model~\cite{Dahmen09,Dahmen11} match our results for the 1D center-of-mass conserving swelling models. 
A mapping between the swelling and avalanche models thus provides an alternative derivation of the scaling in eqn.~\ref{rhon}. 
(Note that while the exponents are universal, the prefactors are model specific~\cite{Dahmen09}.) 
Going the other direction, our analysis of cluster sizes that is based on a Brownian bridge (for $\gamma \rightarrow 1$) and initial overlaps ($\gamma \rightarrow 0$) could provide fresh insight into slip avalanches.

\textit{Generating clusters of a given size.}--- 
In the following two sections, we will address the relaxation within a cluster of size $n$, namely, $\fmovn$ versus time. 
In order to study these systems numerically, we have developed an algorithm to randomly generate the positions of $n$ particles so that their final state is a single isolated cluster of size $n$. 

We start by placing $n$ particles at random on a line of length $\gamma n$ with periodic boundary conditions. 
If these particles are evolved under swelling to size $\gamma$ (in the limit $\epsilon \rightarrow 0$), all the particles will merge into a single cluster that wraps around the entire system, so that there are no edges to the cluster. 
To avoid this, we swell the particles to size $\gamma - \delta$ for some $\delta>0$. 
If $\delta$ is too large, the system will break up into multiple clusters. 
We thus test smaller and smaller $\delta$, until the end result is a single cluster with a gap between only one pair of particles. 
Returning to the initial configuration, we ``cut" the system between these two particles, so that they are explicitly at the ends of the cluster. 
Generating many clusters of exactly size $n$ in this way, we can study the average properties of $\fmovn (t)$.

\section{Purely-repulsive model}
\label{repulsive}

In this section, we will derive the behavior of $\fmovn$ versus time in the purely-repulsive model (Fig.~\ref{Fig1}d), where particles receive only repulsive kicks. 
We will then use it to predict the behavior of $\fmov (t)$.

\textit{Relaxation within a cluster.}---
A natural unit of time is given by considering two overlapping particles in isolation, which will take $\sim \gamma/\epsilon$ cycles to separate. 
Hereafter for the purely-repulsive model, we rescale times by $\gamma/\epsilon$ and distances by $\gamma$. 
We find our results are independent of $\epsilon$ in these units, for $\epsilon < 0.01$. 

\begin{figure}[bt]
\centering 
\begin{center} 
\includegraphics[width=0.45\textwidth]{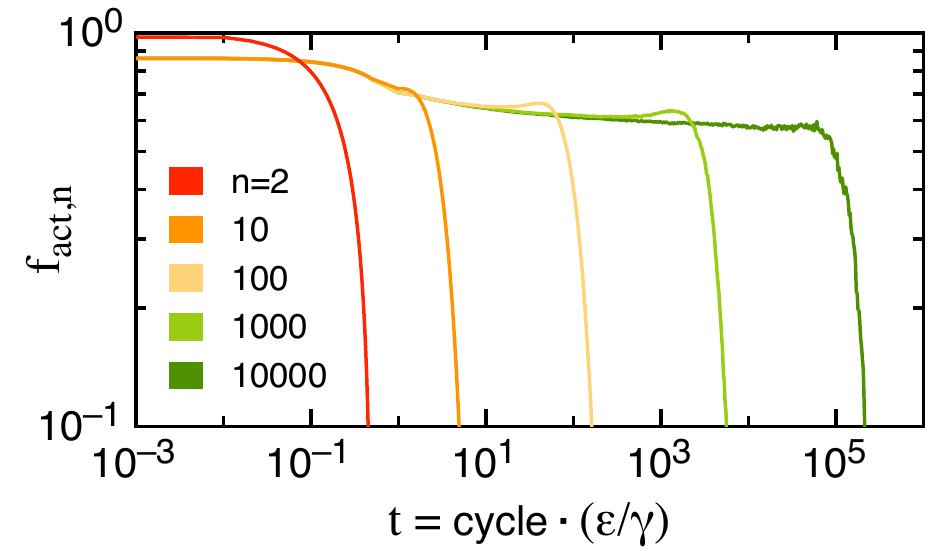} 
\end{center}
\caption{
{\bf Relaxation of clusters of fixed final size for the one-dimensional purely-repulsive model.} 
Fraction of active particles, $\fmovn$, versus time for clusters of different size, $n$. 
For large $n$, the behavior is approximately given by a step function. 
For each $n$, the data are averaged over $(2\times 10^6)/n$ simulations. 
}
\label{Fig3}
\end{figure}

\begin{figure}[bt]
\centering 
\begin{center} 
\includegraphics[width=0.6\textwidth]{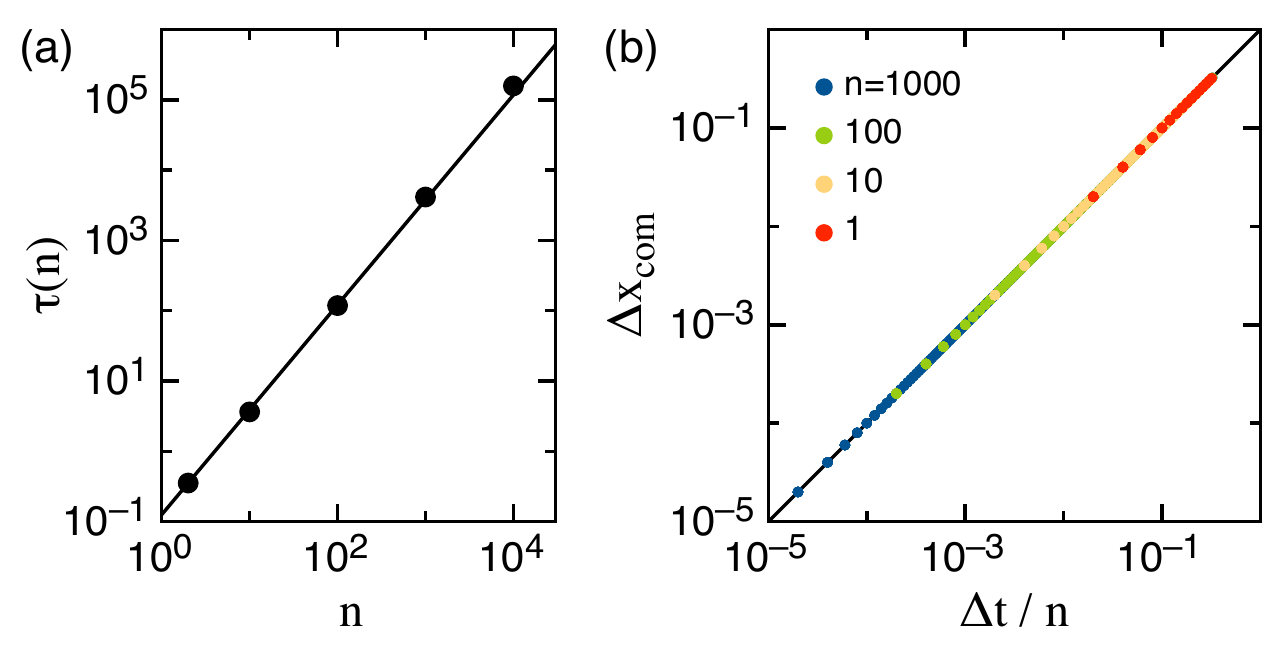} 
\end{center}
\caption{
{\bf Relaxation time for clusters in the purely-repulsive model as a function of size, $n$.} 
{\bf (a)} Relaxation time versus final cluster size, $n$, measured at the time $\fmovn$ falls below $0.29$ (half of the plateau value of $\fmovn$ that we observe for $n=10^4$). 
Solid line: $\tau(n) = 0.126 n^{1.49}$. 
{\bf (b)} Center-of-mass velocity, $\Delta x_{\text{com}}$, of a chain of $n$ particles. 
The data collapse to a rescaled velocity of $1/n$ (solid line). 
Distance is measured in units of $\gamma$ and time is measured in units of cycles $\cdot$ $(\epsilon/\gamma)$. 
}
\label{Fig4}
\end{figure}

Figure~\ref{Fig3} shows $\fmovn (t)$, averaged over many clusters. 
The function approaches a plateau and then falls precipitously to zero at a characteristic time, $\tau(n)$, which grows with $n$. 
We plot the characteristic timescale in Fig.~\ref{Fig4}a. 
This timescale is found to grow as a power law: 
\begin{equation}
\tau(n) = \tau_0 n^{\alpha}, 
\label{taun}
\end{equation}
\noindent where we measure $\tau_0=0.126 \pm 0.003$ and $\alpha=1.49 \pm 0.02$. 

In the remainder of this section, we present an argument that gives $\alpha = 1.5$. 
We start by considering an initial condition where all $n$ particles are evenly spaced, having the same finite overlap with their nearest neighbors. 
On the left side of the cluster, relaxation begins with the leftmost particle moving by $\epsilon$ per cycle. 
Meanwhile, all the particles in the ``bulk" receive canceling kicks and are thus not displaced, but because they are still kicked they are still considered to be active. 
Once the leftmost particle moves off its neighbor, the two leftmost particles are then mobile, and can gradually move to the left on average. 
Relaxation proceeds as a ``chain" of $i$ particles, each with an overlap or gap to the next particle of size $<2\epsilon$, translates to the left. 
Once this chain moves off the bulk, it increases its size to $(i+1)$ particles, and decreases the size of the bulk by $1$ particle. 
(Note that although the center of mass of this chain is moving to the left, another chain on the right-hand-side of this group of $n$ particles is moving to the right, and the center of mass of the entire overlapping group of $n$ particles remains fixed.) 

As shown in Fig.~\ref{Fig4}b, the center of mass of a chain of size $i$ moves at a speed $1/i$. 
Thus, the stage where the leftmost $i$ particles are moving off the bulk will last a duration $i\cdot \text{overlap}(i)$, where $\text{overlap}(i) = 1 - (x_{i+1} - x_i)$. 
Summing over all stages, we get a total relaxation timescale:
\begin{equation}
\tau(n) = \sum_{i=1}^{n/2} i \cdot \text{overlap}(i). 
\label{timesum}
\end{equation}

We now consider the function $\text{overlap}(i)$. 
The random initial particle positions can be decomposed into Fourier modes that capture the density fluctuations on all length-scales. 
Specifically, we decompose the displacements of the initial particle positions from their final positions on an interval of width $n$ that contains the entire cluster. 
For convenience, we use a ``half-range" cosine series, $\{A_k \cos[k \pi (i-0.5)/n]\}$, which is complete and orthogonal on the interval $[0,n]$. 
Here, the particle index, $i$, runs from $1$ to $n$. 
The lowest Fourier mode ($k=1$) compresses the particles towards the center of the cluster when $A_1>0$. 

\begin{figure}[bt]
\centering 
\begin{center} 
\includegraphics[width=0.9\textwidth]{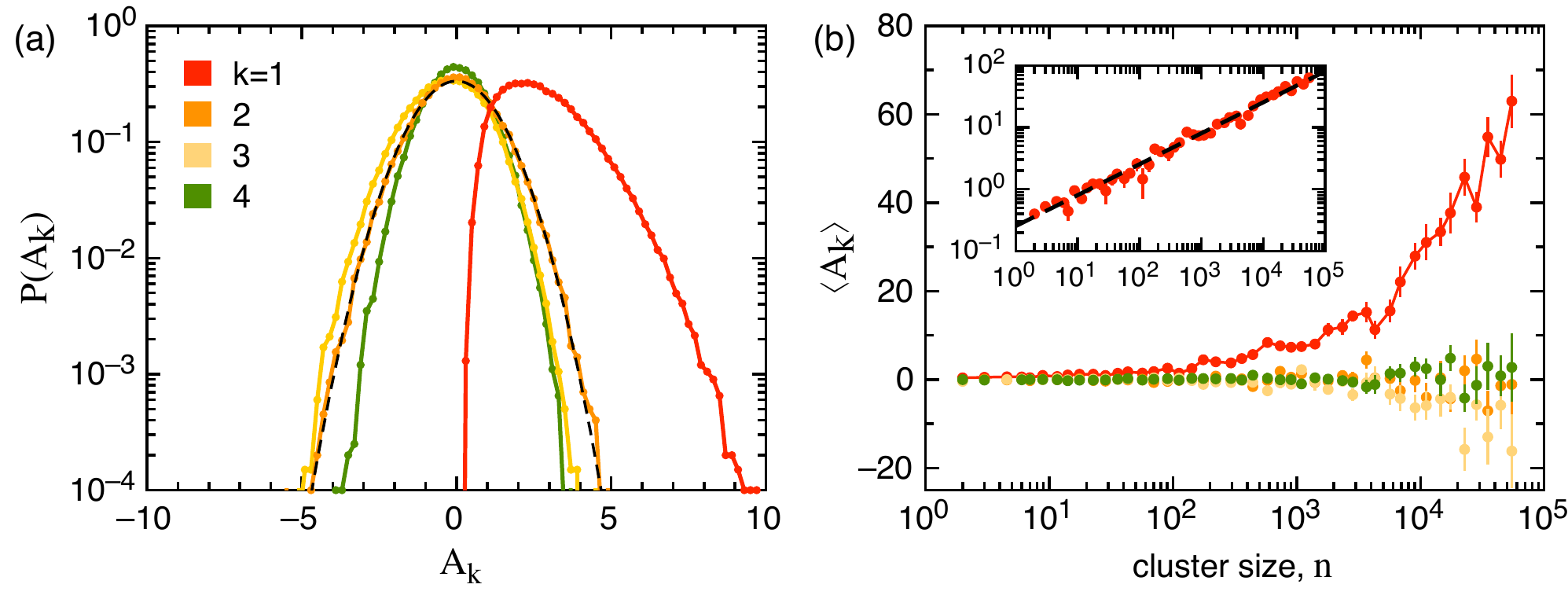} 
\end{center}
\caption{
{\bf Fourier coefficients for initial spatial distribution of particles in the one-dimensional models.} 
{\bf (a)} Probability distribution of the $k^{\text{th}}$ Fourier coefficient, $A_k$, for a cluster of $n=100$ particles (computed by calculating $A_k$ for $10^5$ clusters). 
Only positive values are found for the lowest mode, $A_1$, corresponding to an increase in density in the center of the cluster. 
The distributions for $k \geq 2$ are described well by a Gaussian centered at $0$, as shown by the dashed line for $A_2$. 
{\bf (b)} Mean value of $A_k$ versus cluster size, $n$. 
The lowest mode, $A_1$, steadily increases with $n$, whereas the higher modes fluctuate around zero. 
Inset: coefficient for the lowest mode, $A_1$ (dashed line: $0.254 \sqrt{n}$). 
}
\label{Fig5}
\end{figure}

Figure~\ref{Fig5}a shows the probability distribution of the $A_k$ for the lowest four modes for a cluster of size $n=100$. 
Notably, $A_1$ is always positive, whereas the other coefficients are grouped around zero. 
We can understand this by noting that a negative $A_1$ would correspond to an \textit{expanded} center of the cluster, which would thus not merge into a single cluster over the course of the relaxation. 
Higher modes can however be negative, as they decorate this long-wavelength central compression. 
Thus, the mean of $A_1$ is positive, whereas the mean of higher $A_k$ tend to zero. 

Figure~\ref{Fig5}b shows how this mean, $\langle A_k\rangle$, varies with cluster size, $n$. 
The inset shows that $\langle A_1\rangle$ grows as $\sqrt{n}$. 
This scaling follows from noting that each of the $n$ particles can contribute randomly to the mode, giving it fluctuations of order $\sqrt{n}$. 
Yet the resulting $A_1$ cannot be negative, so the rectified fluctuations have a mean of the same size. 

The timescale for relaxation of a single cluster will be determined by this lowest $k=1$ mode, not only because its coefficient dominates for large $n$, but also because it represents the longest lengthscale over which particles must be transported. 
In this mode, all particles have finite overlap with their neighbors, and the overlap is approximately: $(A_1 \pi /n) \sin[\pi (i-0.5)/n)]$, where $A_1 \approx 0.254 \sqrt{n}$ as shown in Fig.~\ref{Fig5}b. 
Plugging into eqn.~\ref{timesum} and converting to an integral, we find: $\tau(n) \approx 0.081 n^{1.5}$, in good agreement with the data in Fig.~\ref{Fig4}a. 

\begin{figure}[bt]
\centering 
\begin{center} 
\includegraphics[width=0.9\textwidth]{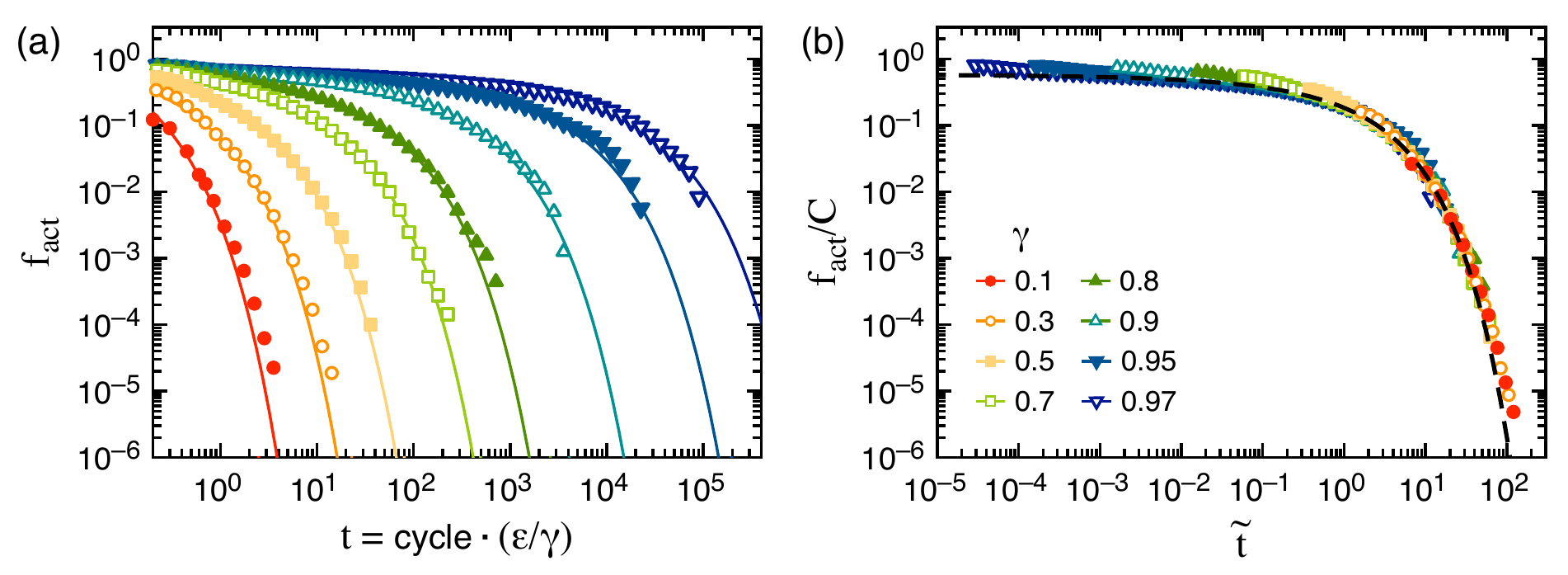} 
\end{center}
\caption{
{\bf Collapse of relaxation curves in the purely-repulsive model.} 
(a) $\fmov$ versus time for a wide range of $\gamma$ from $0.1$ to $0.97$. 
Solid lines: eqn.~\ref{fmov-repuls}. 
(b) The data collapse onto a master curve when the axes are rescaled. 
Rescaled time: $\tilde{t} = (t/\tau_0)(1-\geff)^{2\alpha}$. 
Dashed line: eqn.~\ref{fmov-repuls}. 
}
\label{Fig6}
\end{figure}

\textit{Reconstructing the relaxation dynamics.}---
Harnessing the results thus far, we can now write down a formula for $\fmov(t)$ by using the decomposition in eqn.~\ref{fmovsum}. 
We use eqn.~\ref{rhon} for $\rho(n,\gamma)$, and we find that the $\fmovn(t)$ data are approximated reasonably well by a step function of height $0.58$ and duration $\tau(n)$ given by eqn.~\ref{taun}. 
Plugging into eqn.~\ref{fmovsum} and approximating the sum with an integral, we find the simple form: 
\begin{equation}
\fmov(\tilde{t}) = 0.58 \text{ } C \erfc\left( \tilde{t}^{\frac{1}{2\alpha}} / \sqrt{2}\right),
\label{fmov-repuls}
\end{equation}
\noindent where $\erfc(x)$ is the complementary error function, and we have introduced a rescaled time:
\begin{equation}
\tilde{t} \equiv (t/\tau_0)(1-\geff)^{2\alpha}, 
\label{trescale}
\end{equation} 
\noindent with $\tau_0$ and $\alpha$ as defined in eqn.~\ref{taun}. 
\noindent The only dependance of eqn.~\ref{fmov-repuls} on $\gamma$  is through $C$ and $\tilde{t}$ (the latter via only $\geff$). 

In Fig.~\ref{Fig6}, we show relaxation data for a wide range of $\gamma$. 
The data follow the prediction as a function of time, and they collapse cleanly when plotted in rescaled coordinates, $\fmov/C$ and $\tilde{t}$. 
The only discrepancy is at early times, where the data peel away slightly from the prediction (visible in Fig.~\ref{Fig6}b). 
This is because the $\fmovn$ are higher for $t<1$ (see Fig.~\ref{Fig3}). 
To account for these finer features, one could construct a more detailed approximation to the $\fmov(t)$ data. 
However, our approach is to take the absolute simplest model (i.e., a step function) to highlight how well the full relaxation data can be described with only the basic structure of its components.

\section{Neutral model}
\label{neutral}

In this section, we will derive the behavior of $\fmovn$ versus time in the neutral model (Fig.~\ref{Fig1}c), where particles receive repulsive or attractive kicks with equal probability. 
We will then use it to predict the behavior of $\fmov (t)$. 
We will show that the $\fmovn (t)$ display two distinct timescales, which leads to a wide range of possible stretching exponents as the swelling amplitude, $\gamma$, is varied.

\begin{figure}[bt]
\centering 
\begin{center} 
\includegraphics[width=0.45\textwidth]{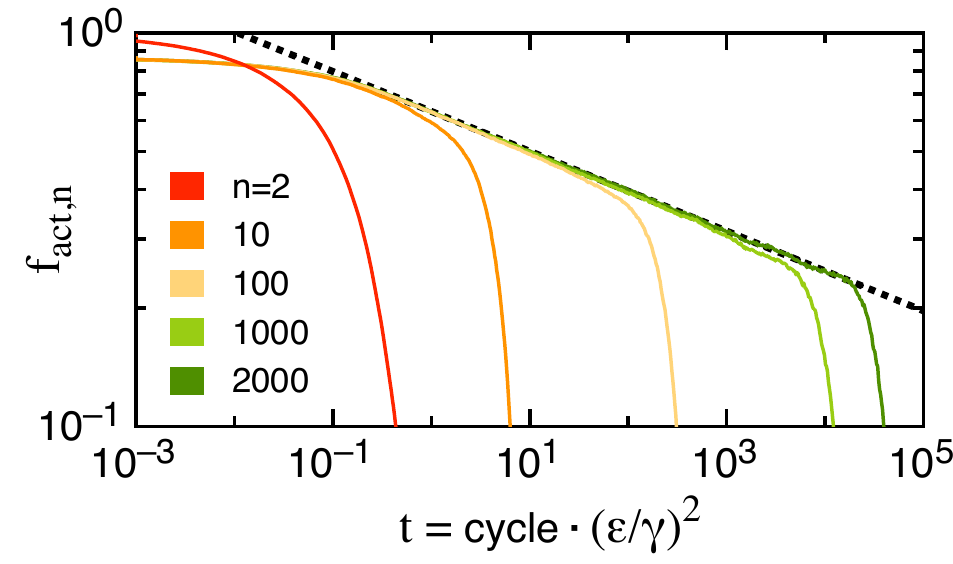} 
\end{center}
\caption{
{\bf Relaxation of clusters of fixed final size in the one-dimensional neutral model.} 
Fraction of active particles, $\fmovn$, versus time. 
For $t \gtrsim 0.2$, the function relaxes as a power-law, $\fmovn(t) \propto t^{-0.101}$, before falling sharply to zero. 
Dashed line: eqn.~\ref{fmovn-repatt}. 
For each $n$, the data are averaged over $10^5/n$ simulations. 
}
\label{Fig7}
\end{figure}

\textit{Relaxation within a cluster.}---
Figure~\ref{Fig7} shows $\fmovn (t)$ for several final cluster sizes, $n$, where in this model we rescale time by $(\gamma/\epsilon)^2$, corresponding to the number of cycles for a particle to diffuse by its diameter. 
The data decay gradually at first and then fall to zero. 
For $t \gtrsim 0.2$ but before this drop-off, we find a good fit to a power law: 
\begin{equation}
\fmovn(t) = 0.63 \text{ } t^{-0.101 \pm 0.002}. 
\label{fmovn-repatt}
\end{equation}

\begin{figure}[bt]
\centering 
\begin{center} 
\includegraphics[width=0.28\textwidth]{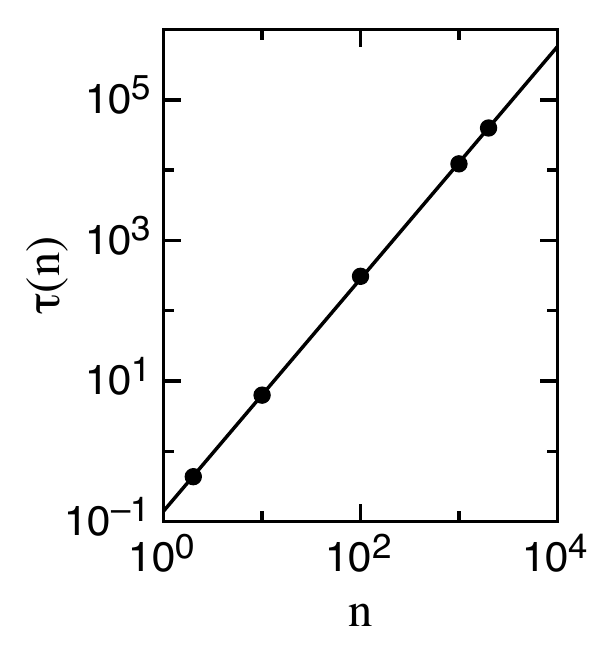} 
\end{center}
\caption{
{\bf Relaxation timescale versus final cluster size in the neutral model.}  
Solid line: fit to the data, $\tau(n) = 0.14 n^{1.65}$. 
}
\label{Fig8}
\end{figure}

As in the previous section, we extract a characteristic timescale, $\tau(n)$, from these curves. 
Here, to isolate the sharp drop-off from the more gradual relaxation that precedes it, we measure the time when $\fmovn (t)$ falls below $0.1$. 
Figure~\ref{Fig8} shows that the $\tau(n)$ data follow a power law (eqn.~\ref{taun}), as in the purely-repulsive case. 
Here we measure $\tau_0=0.14 \pm 0.01$ and $\alpha=1.65 \pm 0.02$. 
(We note that in the repulsive-attractive case, the same decomposition of the initial particle positions into Fourier modes applies. 
However, estimating a timescale for relaxation within a single Fourier mode is complicated by the fact that particles can be kicked towards each other in this model.) 

\textit{Reconstructing the relaxation dynamics.}---
We now reconstruct $\fmov(t)$ by plugging these results into the original decomposition, eqn.~\ref{fmovsum}. 
Because the power-law decay in $\fmovn(t)$ is the same for all values of $\gamma$, we can simply factor this term out of the sum. 
The remaining sum can be analyzed in the same manner as in the previous section. 
Thus, we find $\fmov (t)$ to follow eqn.~\ref{fmov-repuls} with an additional factor that is given by eqn.~\ref{fmovn-repatt}. 
In particular, 
\begin{equation}
\fmov (t,\tilde{t}) = 0.63 \text{ } C \text{ } t^{-0.101} \erfc\left( \tilde{t}^{\frac{1}{2\alpha}} / \sqrt{2}\right). 
\label{fmov-repatt}
\end{equation}
\noindent Notably, $\fmov$ depends on both $t$ and $\tilde{t}$ in this model variant. 
The existence of two distinct timescales comes from the fact that the initial power-law relaxation of a cluster is independent of its size, $n$, whereas the total lifetime of a cluster depends strongly on $n$. 

\begin{figure}[bt]
\centering 
\begin{center} 
\includegraphics[width=0.85\textwidth]{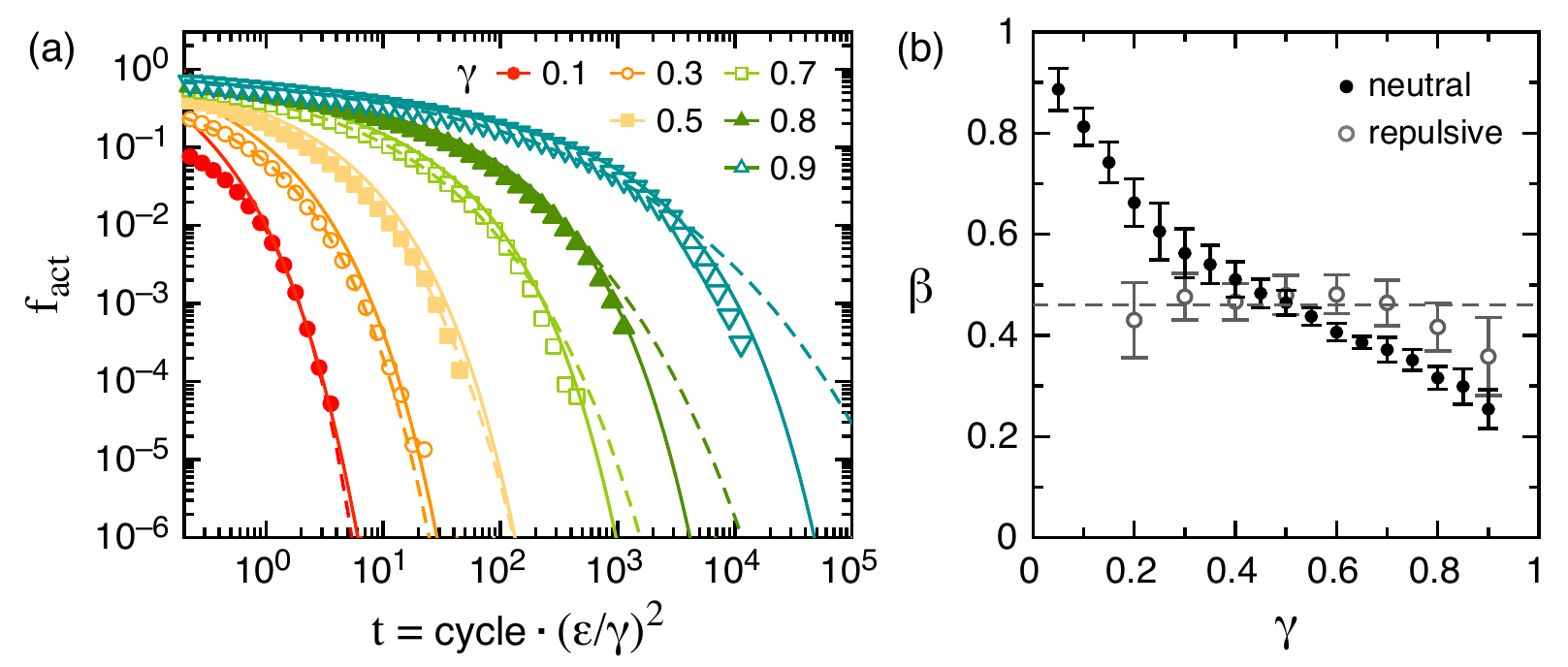} 
\end{center}
\caption{
{\bf Realizing different values of the stretching exponent.} 
{\bf (a)} $\fmov$ versus time in the neutral model. 
Dashed lines: stretched-exponential fits. 
Solid lines: eqn.~\ref{fmov-repatt}. 
{\bf (b)} Best-fit stretching exponent versus swelling size, $\gamma$. 
In the purely-repulsive model, $\beta$ is not a strong function of $\gamma$. 
Dashed line: constant fit to repulsive data, $\beta = 0.46$. 
In the neutral model, $\beta$ approaches $1$ for small $\gamma$, and continuously drops to values as low as $0.25$ for $\gamma=0.9$. 
}
\label{Fig9}
\end{figure}

Figure~\ref{Fig9}a shows our prediction for $\fmov$ at several values of $\gamma$, which describe the data well. 
(Once again, we note that the curves peel off the data at early times -- in this case our power-law approximation of $\fmovn$ overshoots the data for small $t$, especially for small clusters, as shown in Fig.~\ref{Fig7}.) 

We also fit the data at each swelling amplitude to a stretched-exponential. 
For small and intermediate $\gamma$, these fits describe the data very well. 
For $\gamma=0.8$ and $0.9$, the fits do well at early times, but they are clearly not good at the tails. 
Yet, even when we have $\beta$ as small as $0.25$ (for $\gamma=0.9$), our functional form, eqn.~\ref{fmov-repatt}, captures the data remarkably well, spanning more than $4.5$ decades in time. 

Figure~\ref{Fig9}b shows the values of the stretching exponent, $\beta$, obtained from our fits. 
We find that $\beta$ varies over a wide range in the neutral model, approaching $1$ for small $\gamma$ and continuously dropping down to $0.25$ for $\gamma=0.9$. 
In contrast, $\beta$ is not a strong function of $\gamma$ in the purely repulsive model. 
Thus, we see that the crucial difference between the relaxation in the two models stems from the form of the relaxation within an individual cluster -- in the neutral case, $\fmovn(t)$ possesses a power-law decay, whereas in the purely-repulsive model, $\fmovn(t)$ is approximately constant. 
In supercooled liquids, the stretching exponent decreases as temperature is lowered~\cite{Cavagna09}. 
Our neutral-interaction particle model exhibits this nontrivial behavior simply by increasing the swelling size, $\gamma$. 
The extreme stretching is associated with larger clusters of interacting particles that must relax. 

At the critical amplitude that marks the transition to irreversibility (i.e., $\gamma=\gamma_{\text{c}}$ \cite{Corte08}), one expects power-law relaxation, as was observed in a 2D version of the model \cite{Schrenk15}. 
Investigating this limit in our neutral model, we find that eqn.~\ref{fmov-repatt} approaches $\fmov (t) = 0.63 \text{ } C \text{ } t^{-0.101}$ as $\gamma \rightarrow 1$.


\section{Conclusion}

Our results provide a concrete example of an extremely simple set of dynamics that produces a wide range of stretching exponents. 
We have shown that randomness in the initial particle positions is sufficient to produce stretched-exponential relaxation, even when the collision rule is completely deterministic. 
Allowing the collisions to be repulsive or attractive with equal probability, a wide range of stretching exponents could be obtained, depending on the swelling size. 

Both one-dimensional model variants studied here are well fit by stretched-exponential relaxation (eqn.~\ref{strexp}), but we have derived alternative functional forms for these particular models. 
Thus, for our purposes eqn.~\ref{strexp} may be viewed as a convenient two-parameter fitting function~\cite{Menon92,Apitz05}. 

In both model variants, density fluctuations in the initial state lead to a broad distribution of cluster sizes, which was a key ingredient in understanding the dynamics of the two models. 
However, the value of the power-law exponent $\alpha$ describing the lifetime of a cluster of size $n$ (eqn.~\ref{taun}) was not particularly important for obtaining long tails in the relaxation; apparently $\alpha$ does not have to be fine tuned to yield approximately stretched-exponential behavior. 

More broadly, we have found that this simple model that was introduced for a non-Brownian suspension looks to be a good model for glassy dynamics. 
Indeed, several signatures of glassy behaviors have recently been reported in a two-dimensional version of the model~\cite{Tjhung15,Tjhung16}. 
Here we have shown that both two-dimensional and one-dimensional versions of the model exhibit glassy relaxations. 
We expect this is because the model captures fluctuations in density, which would clearly be there in a glass. 
Future work could study relaxation in similar models of sheared amorphous solids where the density is much larger~\cite{Regev13,Fiocco14}. 
In the simple model variants we studied, particles interact because of randomness, and they must get out of each other's way to relax. 
This disorder is what gives rise to the slow relaxation. 
This remarkably simple set of rules is just sufficient to generate this nontrivial behavior, and is therefore a promising avenue for studying glassy systems.

\begin{acknowledgments} 
We are grateful to Leo Kadanoff, our teacher, mentor, and friend, for many years of scientific discussions. 
He had the ability to understand complex interacting systems by isolating the common simple ingredients. 
This has been an inspiration to all of us in the soft-matter physics community. 
We thank Thomas A. Caswell for assistance optimizing the simulations, and Karin Dahmen and Daniel Hexner for stimulating conversations.  
This work was supported by NSF Grant DMR-1404841 and by NSF MRSEC DMR-1420709. 
J.D.P. acknowledges the Donors of the American Chemical Society Petroleum Research Fund for partial support of this research. 
Use of computation facilities funded by the US Department of Energy, Office of Basic Energy Sciences, Division of Materials Sciences and Engineering, Award No.~DE-FG02-03ER46088 is gratefully acknowledged. 
\end{acknowledgments}


%

\end{document}